\documentstyle[12pt]{article}
\newcommand{\beq}{\begin{equation}}
\newcommand{\eeq}{\end{equation}}
\newcommand{\bea}{\begin{eqnarray}}
\newcommand{\eea}{\end{eqnarray}}

\begin{document}
\setcounter{page}{0}
\topmargin 0pt
\oddsidemargin 5mm
\renewcommand{\thefootnote}{\fnsymbol{footnote}}
\newpage
\setcounter{page}{0}
\begin{titlepage}
\begin{flushright}
QMW 93-25
\end{flushright}
\begin{flushright}
hep-th/9310014
\end{flushright}
\vspace{0.5cm}
\begin{center}
{\large {\bf The WZNW Model By A Perturbation Of
Witten's Conformal Solution}} \\
\vspace{1.8cm}
\vspace{0.5cm}
{\large Oleg A. Soloviev
\footnote{e-mail: soloviev@V1.PH.QMW.ac.uk}\footnote{Work supported by
S.E.R.C.}}\\
\vspace{0.5cm}
{\em Physics Department, Queen Mary and Westfield College, \\
Mile End Road, London E1 4NS, United Kingdom}\\
\vspace{0.5cm}
\renewcommand{\thefootnote}{\arabic{footnote}}
\setcounter{footnote}{0}
\begin{abstract}
{The WZNW model at the Witten conformal point is perturbed by the
$\sigma$-model term. It is shown that in the large level $k$ limit the
perturbed WZNW system with negative $k$ arrives at the WZNW model with positive
$k$.}
\end{abstract}
\vspace{0.5cm}
\centerline{September 1993}
 \end{center}
\end{titlepage}
\newpage
The Wess-Zumino-Novikov-Witten (WZNW) model [1-4] is a $\sigma$-model extended
by the Wess-Zumino term [5]. The action of the theory is written as follows [2]
\begin{equation}
S=(1/4\lambda^2)\int d^2x Tr (\partial_\mu g\partial^\mu g^{-1})\;+\;k\Gamma,
\end{equation}
where $g$ is the matrix field taking its values on the Lie group $G$; $k$ is
the level of the affine Lie algebra $\hat{\cal G}$ associated with the Lie
algebra ${\cal G}$ of the given Lie group $G$; $\lambda$ is a $\sigma$-model
coupling constant; $\Gamma$ is the Wess-Zumino term [5]. For compact groups the
Wess-Zumino term
is well
defined only modulo $2\pi$ [2], therefore, the parameter $k$ must be an integer
in order for the quantum theory to be single valued with the multivalued
classical action. For noncompact groups the level may be arbitrary.

The remarkable property of the WZNW model, due to Witten, is that at the
particular value of its coupling constant
\begin{equation}
\lambda^2=|4\pi/k|
\end{equation}
both the affine and conformal symmetries of the classical theory are promoted
to the quantum level, making the system be exactly soluble [2-4]. It is
interesting that Witten has found the
exact value of the fixed point from the one
loop renormalization beta function [2]. To be sure that it was not a miracle
the two loop calculations have been done as well [6]. They demonstrated that
Witten's solution remains exact up to two loops within the path integral
formalism. Moreover, at the Witten conformal point, the theory can be quantized
nonperturbatively either within the algebraic current-current approach [3] or
by
means of the free field representation method [7].

The aim of the present paper is to display that the WZNW model with negative
$k$ flows to the WZNW model with positive $k$.

Let us consider the theory with the following action
\begin{equation}
S=S^{*}\;+\;\epsilon\int d^2z(k^2/4) Tr(\partial g\cdot\bar\partial g^{-1}),
\end{equation}
where $S^{*}$ is the WZNW action at the Witten conformal point; $\epsilon$ is a
small parameter. For our purposes it is more convenient to use complex
coordinates. For example, the action $S^{*}$ in eq. (3) in these coordinates
takes the form
\begin{equation}
S^{*}=-(k/4\pi)\{(1/2)\int d^2z Tr |g^{-1}dg|^2\;+\;(i/3)\int
d^{-1}Tr(g^{-1}dg)^3\}.\end{equation}
Note that at the classical level the action in eq. (3) follows directly from
eq.
(1) by setting the coupling constant $1/\lambda^2$ to the specific value
$(k/8\pi)+\epsilon k^2/4$.

At the quantum level the WZNW model $S^{*}$ is an exact conformal
theory. Therefore, the second term in eq. (3) appears to be a certain
perturbation to the conformal system $S^{*}$. To first order in
$\epsilon$, we can write
\begin{equation}
S-S^{*}=-\epsilon \int d^2z :\phi^{a\bar a}\cdot J_a\cdot\bar J_{\bar a}:,
\end{equation}
where in the r.h.s we have changed the classical $\sigma$-model perturbation
by a normal ordered product of the following three operators
\begin{eqnarray}
\phi^{a\bar a}&=&Tr(g^{-1}t^agt^{\bar a}),\nonumber\\
J_a&=&(1/2)\eta_{ab}Tr( kg^{-1}\partial g t^b),\\
\bar J_{\bar a}&=&(1/2)\eta_{\bar a\bar b}Tr(k\bar\partial gg^{-1}t^{\bar
b}).\nonumber\end{eqnarray}
Apparently, in the leading order in $\epsilon$ we can define normal ordering in
eq. (5) with respect to the conformal model $S^{*}$ by the rule
\begin{equation}
O(z,\bar z)\equiv :\phi^{a\bar a}\cdot J_a\cdot\bar J_{\bar a}:(z,\bar z) =
\oint{dw\over 2\pi}\oint{d\bar w\over 2\pi}{J_a(w)\cdot \bar J_{\bar a}(\bar
w)\cdot\phi^{a\bar a}(z,\bar z)\over |z-w|^2}.\end{equation}
In the numerator of eq. (7) the product is understood as an operator
product expansion (OPE).
In the theory $S^{*}$ the operators $\phi^{a\bar a}$ in (6) are
primary operators. Therefore, all correlation functions built of the operator
$O$ can be computed by using the Ward identities [3, 8]. Of
course, such normal ordering can get corrected by terms of higher order in
$\epsilon$. With respect to $S^{*}$ the operators in eq. (6) have the
following conformal dimensions
\begin{eqnarray}
(\Delta_\phi,\bar\Delta_\phi)&=&\left({c_2\over c_2+k},{c_2\over c_2+k}\right),
\nonumber\\
(\Delta_J,\bar\Delta_J)&=&(1,0),\\
(\Delta_{\bar J},\bar\Delta_{\bar J})&=&(0,1),\nonumber\end{eqnarray}
with $c_2$ the quadratic Casimir operator eigenvalue in the adjoint
representation of the Lie algebra ${\cal G}$.

Let us turn now to the large $|k|$ limit. It is not hard to see that in this
limit the perturbation operator becomes a quasimarginal operator with
anomalous dimension
\begin{equation}
(\Delta,\bar\Delta)=(1+c_2/k,1+c_2/k).\end{equation}
Hence, when $k$ is negative - that is meaningful for noncompact groups - the
given perturbation is to be classified as
relevant; whereas for positive $k$ one should refer to  an irrelevant
perturbation.

Given the perturbation one can try to calculate the renormalization beta
function associated with the coupling $\epsilon$. Away of criticality, where
$\epsilon\ne0$, the beta function is defined according to [9-11]
\begin{equation}
\beta=[(2-(\Delta+\bar\Delta))\epsilon\;-\;\pi C\cdot\epsilon^2\;+\;{\cal
O}(\epsilon^3)],
\end{equation}
where $(\Delta,\bar\Delta)$ are given by eq. (9). The $C$ is taken here to be
the coefficient of the three point function
\begin{equation}
\langle O(z_1)\;O(z_2)\;O(z_3)\rangle={C\over|z_{12}|^{\Delta+\bar\Delta}
|z_{13}|^{\Delta+\bar\Delta}|z_{23}|^{\Delta+\bar\Delta}}\end{equation}
when the two point functions are normalized to unity.

Bearing in mind the definition of $O$ we obtain the following expression for
the coefficient $C$
\begin{equation}
C\,=\,C_{\phi\phi\phi}^{a\bar a\;b\bar b\;c\bar
c}C_{JJJ}^{abc}C_{\bar J\bar J\bar J}^{\bar a\bar b\bar c}.\end{equation}
where, respectively, the coefficient $C_{JJJ}$ and $C_{\bar J\bar J\bar J}$
coincide with the coefficient of the three point functions of the currents $J$
and $\bar J$; whereas $C_{\phi\phi\phi}$ is the coefficient of the three point
function of the operator $\phi$.
The coefficients $C_{JJJ}$ and $C_{\bar J\bar J\bar J}$ can be easily extracted
from the affine current algebra
\begin{equation}
C_{JJJ}^{bcd}=if^{bcd},\;\;\;\;\;\;\;C_{\bar J\bar J\bar J}^{\bar
b\bar c\bar d}=if^{\bar b\bar c\bar d}.\end{equation}
The coefficient $C_{\phi\phi\phi}$ in its turn is
found from the four point function [12]
\begin{eqnarray}
\langle\phi(\infty)\;\phi(1)\;g^{-1}(z)\;g(0)\rangle,\nonumber
\end{eqnarray}
by looking at the most singular term in this function as $z\rightarrow0$.
In ref. 12 the coefficient $C_{\phi\phi\phi}$ is
calculated in powers of $1/N$ for $G=SU(N)$. In the large $|k|$ limit
we can argue that the coefficient $C_{\phi\phi\phi}$ is given by
\begin{eqnarray}
C_{\phi\phi\phi}^{a\bar a\;c\bar c\;d\bar d}&=&(1/c_2)
\{Tr(t^a[t^c,t^d]_-)\cdot Tr(t^{\bar a}[t^{\bar c},t^{\bar d}]_-)\nonumber
\\ & & \\
&+&Tr(t^a[t^c,t^d]_+)\cdot Tr(t^{\bar a}[t^{\bar c},t^{\bar d}]_+)\}.
\nonumber\end{eqnarray}

Combining formulas (13), (14) in eq. (12) we get $C$ in the form
\begin{equation}
C= c_2.\end{equation}
Note that the coefficient $C$ in eq. (15) corresponds to the operator $O$
normalized to one so that all coefficients like $(\dim G)^2$ are dropped out.

With the given expression for $C$ the equation for the beta function takes the
form
\begin{equation}
\beta={-2c_2\over k}\epsilon\;-\;\pi c_2\epsilon^2\;+\;{\cal
O}(\epsilon^3).\end{equation}
One can easily solve this equation to find fixed
points of the beta function. There are two solutions
\begin{equation}
\epsilon_1=0,\;\;\;\;\;\;\;\;\;\epsilon_2=-2/(\pi k).\end{equation}
The first one is nothing but Witten's conformal point of the WZNW
model with negative $k$; whereas the second solution signifies a
conformal point in the WZNW model with positive $k$. Indeed the $\sigma$-model
coupling constant at the second conformal point is
\begin{equation}
\lambda^{*}_2=-4\pi/k=4\pi/|k|.
\end{equation}
It is well known that the WZNW model with the given coupling constant is an
exact theory [2].

So, the expansion of the WZNW model around Witten's critical
point allowed us to discover a flow between conformal solutions with positive
and negative $k$. It might be very interesting to check what is going on in
higher orders of the perturbation theory because the value of $\lambda$ in eq.
(18) corresponds to lowest order in $1/k$.

We can compute also the Virasoro central charge at the second fixed point by
using Cardy-Ludwig formula [10]
\begin{equation}
c(\epsilon_2)=c(\epsilon_1)\;+\;{8c_2\over k^3}.
\end{equation}
So, when $k$ is negative - that corresponds to the case of the relevant
perturbation - the central charge becomes lower along the flow from Witten's
conformal fixed point to the second nontrivial conformal point. Such behavior
one might expect for unitary flows according to the Zamolodchikov's c-theorem
[13].

\par \noindent
{\em Acknowledgement}: I would like to thank G. Mussardo for fruitful
conversations. I thank also J. M. Figueroa-O'Farrill for careful reading of the
manuscript.

\end{document}